\documentclass[manuscript]{aastex}

\usepackage{natbib, aas_macros}
\usepackage{amsmath}
\usepackage{color}
\usepackage{ulem}

\definecolor{MyDarkBlue}{rgb}{0,0.08,0.45}
\definecolor{MyDarkRed}{rgb}{0.8,0.1,0.08}
\definecolor{Red}{rgb}{1.0,0.0,0.2}
\definecolor{Blue}{rgb}{0,0.08,0.95}
\definecolor{LightGrey}{rgb}{0.7,0.7,0.7}

\newcommand{\bs}{\boldsymbol}

\begin{document}
\title{Breaking Taylor-Proudman balance by magnetic field in stellar
convection zone}
\author{H. Hotta}
\affil{Department of Physics, Graduate School of Science, Chiba
university, 1-33 Yayoi-cho, Inage-ku, Chiba, 263-8522, Japan
}

   \begin{abstract}
    We carry out high-resolution calculations for the stellar
    convection zone. The main purpose of this study is to investigate the
    effect of a small-scale dynamo on the differential
    rotation. The solar differential rotation deviates from the
    Taylor-Proudman state in which the angular velocity does not
    change along the rotational axis. To break the
    Taylor-Proudman state deep in the convection zone, it is thought
    that a latitudinal entropy
    gradient is required.
    In this study, we find
    that the small-scale dynamo has three roles in the deviation of the stellar
    differential rotation from the Taylor-Proudman state.
    1.
    The shear of the angular velocity is suppressed. This leads to a situation where the
    latitudinal entropy gradient
    efficiently breaks the Taylor-Proudman state. 2. The perturbation of
    the entropy is increased with suppressing the
    turbulent velocity between
    up- and
    downflows. 3. The convection velocity is reduced. This increases the
    effect of the rotation on the convection. The second and third factors
    increase the latitudinal entropy gradient and break the
    Taylor-Proudman state. We find that the efficient small-scale
    dynamo has a significant impact on the stellar differential rotation.
   \end{abstract}
  
  \clearpage
 \section{Introduction}
 The Sun is rotating differentially. The structure of the differential
 rotation is one of the most important factors in the generation of the
 magnetic field, i.e. the dynamo, because the differential rotation stretches the
 large-scale poloidal field to the toroidal field \citep{1955ApJ...122..293P}.
 The detailed profile of the solar differential rotation has been
 revealed by global helioseismology
 \citep[e.g.][]{1998ApJ...505..390S}. The profile shows the three
 features that deviate from the Taylor-Proudman state, where the angular
 velocity has a cylindrical profile;
 i.e., $\partial\langle\Omega\rangle/\partial z=0$, where $\Omega$
 and $z$ are the angular velocity and the direction of the rotation
 axis, respectively. The angle brackets $\langle\rangle$ shows the longitudinal
 average. The tachocline, the
 near surface shear layer, and the conical profile of the angular
 velocity in the bulk of the convection zone are the deviations. The
 Taylor-Proudman theorem
 is derived from the longitudinal component of the vorticity equation
 as \citep[see the detail in][]{2005ApJ...622.1320R,2011ApJ...740...12H}:
 \begin{eqnarray}
  \frac{\partial \langle\omega_\phi\rangle}{\partial t} =
   \left(\nabla\times\langle{\bs v}\times {\bs
		\omega}\rangle\right)_\phi
   +\lambda\frac{\partial \langle\Omega\rangle^2}{\partial z}
   - \frac{g}{c_\mathrm{p}r}\frac{\partial \langle s\rangle}{\partial
   \theta}
   +\left(
     \nabla\times
     \left\langle\frac{1}{4\pi\rho}(\nabla\times{\bs B})\times{\bs B}\right\rangle
	    \right)_\phi,\label{tp}
 \end{eqnarray}
 where $r$, $\theta$, and $\phi$ show the radius, the colatitude, and
 the longitude in the spherical geometry. $\lambda=r\sin\theta$ is the
 distance from the rotational axis.
 $\rho$,
 ${\bs v}$, $\bs{\omega}$, $\Omega$, $s$, $g$, $c_\mathrm{p}$ and $\bs{B}$
 show the density, the fluid velocity
 in the rotating frame, the vorticity, the angular velocity,
 the specific entropy, the gravitational acceleration, the heat capacity
 at constant pressure, and the magnetic field, respectively.
 When all the other terms in eq. (\ref{tp}) are
 ignored, the Taylor-Proudman
 theorem $\partial\langle\Omega\rangle/\partial z = 0$ is derived.
 %In this discussion, we assume the solar differential rotation is statistically steady
 %($\partial \langle\omega_\phi\rangle/\partial t=0$).
 The deviation of the Taylor-Proudman state
 should be explained with the transport term
 $\nabla\times\langle {\bs v}\times{\bs \omega}\rangle$, the
 baroclinic term (latitudinal entropy gradient)
 $-g/(c_\mathrm{p}r)\partial \langle s\rangle/\partial \theta$, or the
 Lorentz force term.
 \cite{2015ApJ...798...51H} suggest that the near surface shear layer is
 maintained by the transport term as a consequence of sheared
 meridional flow. The tachocline and the conical profile of the
 differential rotation are thought to be maintained by the baroclinic
 term
 \citep{2005ApJ...622.1320R,2006ApJ...641..618M,2011ApJ...742...79B}.
 \cite{2005ApJ...622.1320R} adopted the mean field model, i.e. the turbulence was
 not solved, and suggested that the interaction of the meridional flow and the
 subadiabatic overshoot region can maintain the latitudinal entropy
 gradient, and the Taylor-Proudman state is broken. In the three-dimensional (3D)
 model,
 \cite{2006ApJ...641..618M} put the latitudinal entropy gradient at the
 bottom boundary and maintain the conical profile of the differential
 rotation. \cite{2011ApJ...742...79B} include the radiation zone and
 maintain the large entropy gradient in the overshoot region
 self-consistently, and as a result, the tachocline is also generated.
\par
 On the other hand, it has been reported that the convection zone itself
 has the ability to make the latitudinal entropy gradient
 \citep{2000ApJ...532..593M}. The cool downflow is bent to the equator by
 the Coriolis
 force, and the hot upflow is bent to the pole. As a result, the anisotropy of the
 thermal convection is able to form a negative latitudinal entropy
 gradient. We call this the CZ effect in this
 paper. \cite{2000ApJ...532..593M} reported that this effect is not
 enough to explain the observed differential rotation; i.e., the
 generated entropy gradient is too small. In this letter, we further explore
 the possibility of the CZ effect in 3D convection
 calculations.\par
 One possibility for amplifying the CZ effect is the magnetic
 field. \cite{2015ApJ...803...42H,2016Sci....351..1427} show that in high-resolution calculations, the
 magnetic field is on an equipartition
 level with the kinetic energy of turbulent flow and significantly modifies
 the convective structure. In particular,
 the entropy structure is modified, because the small-scale
 turbulent velocity between
 the up- and downflows is suppressed by the small-scale magnetic
 field. This has the possibility of changing the CZ effect for the latitudinal
 entropy gradient. \cite{2016Sci....351..1427} have already shown that the
 high efficiency of the small-scale dynamo changes the distribution of
 the differential rotation (see their Figure 2 in the
 Supplementary Material). When the small-scale dynamo is
 very effective, the differential rotation tends to become a
 non-Taylor-Proudman state.
 A similar feature is seen in the high magnetic Prandtl number (low magnetic
 diffusivity) calculation in \cite{2017A&A...599A...4K}.
 As \cite{2016Sci....351..1427} use the solar rotation rate, they adopt
 a very large thermal conduction
 on the entropy to avoid a high convection velocity and the resulting polar
 acceleration. Recently, a large number of studies have argued that the high-
 resolution calculations with
 the solar rotation rate tend
 to reproduce the high convection velocity and to cause polar acceleration of the
 differential rotation, which is not consistent with the real Sun
 \citep[e.g.][]{2014MNRAS.438L..76G,2014A&A...570A..43K,2014ApJ...789...35F,2015ApJ...798...51H}.
 To avoid this situation, \cite{2016Sci....351..1427} added a very
 large thermal conductivity to suppress the convection
 velocity. Meanwhile, the small-scale entropy feature is also
 suppressed in the calculation, and it becomes difficult to discuss the
 role of the small-scale dynamo in the creation of the latitudinal
 entropy gradient. In this paper, we decided to change the rotation rate
 to $3\Omega_\odot$ and to exclude the large thermal conduction on the
 entropy. Then we achieve an efficient small-scale dynamo in a relatively
 high-resolution calculation and discuss the importance of this
 small-scale dynamo on the latitudinal entropy gradient. We note that
 there will still be discussion about the small-scale dynamo in a low magnetic
 Prandtl number $\mathrm{P_m}=\nu/\eta$, where $\nu$ and $\eta$ are the
 kinematic viscosity and the magnetic diffusivity
 \citep{2018arXiv180209607K}. The low magnetic 
 Prandtl number is achieved in the solar convection zone.
 This is beyond the scope of this study. Here
 we discuss the small-scale dynamo with $\mathrm{P_m}=1$, which may
 be achieved only with numerical diffusivities.
 
 \section{Model}
 We solve the 3D magnetohydrodynamics (MHD) equations in the
 spherical geometry $(r,\theta,\phi)$. The equations solved in this
 study are listed as:
 \begin{eqnarray}
  \frac{\partial\rho_1}{\partial t} &=& -\frac{1}{\xi^2}\nabla\cdot(\rho
   \bs{v}),\\
  \rho\frac{\partial {\bs v}}{\partial t} &=& -\rho
   \left({\bs v}\cdot\nabla\right){\bs v} - \rho_1 g {\bs e}_r + \nabla
   p_1 +  2\rho{\bs v}\times{\bs
   \Omega}_0+\frac{1}{4\pi}\left(\nabla\times{\bs B}\right)\times{\bs
   B},\\
  \frac{\partial {\bs B}}{\partial t} &=& \nabla\times
   \left({\bs v}\times{\bs B}\right),\\
  \rho T\frac{\partial s_1}{\partial t} &=& -\rho T
   \left({\bs v}\cdot\nabla\right)s + Q_\mathrm{rad},\\
  p_1 &=&
   \left(
    \frac{\partial p}{\partial \rho}
		   \right)_s \rho_1
   +
   \left(
    \frac{\partial p}{\partial s}
   \right)_\rho s_1,
 \end{eqnarray}
 where $p$, and $T$ are the gas pressure and the temperature,
 respectively; the subscript 0 and 1 are the steady background and
 perturbation quantities, respectively;
 $\xi$ is the factor for the reduced speed of sound
 technique \cite[RSST;][]{2012A&A...539A..30H}. We make the speed of
 sound uniform as:
 \begin{eqnarray}
  \xi (r) &=& \xi_0 \frac{c_\mathrm{s}(r)}{c_\mathrm{s}(r_\mathrm{min})},\\
  c_\mathrm{s} &=&
   \sqrt{
   \left(
    \frac{\partial p}{\partial \rho}
	    \right)_s
   },
 \end{eqnarray}
 where $r_\mathrm{min}$ and $c_\mathrm{s}$ are the location of the
 bottom boundary and the adiabatic speed of sound, respectively. We
 adopt $\xi_0=100$. 
 The background density $\rho_0$,
 temperature $T_0$, and the pressure $p_0$ are calculated using Model S
 \citep{1996Sci...272.1286C,2014ApJ...786...24H}. The factors for the
 linearized equation of state, $(\partial p/\partial \rho)_s$ and
 $(\partial p/\partial s)_\rho$, are calculated with the OPAL repository
 \citep{1996ApJ...456..902R}.
 $Q_\mathrm{rad}$ includes the radiative heating around the base of the
 convection zone and the artificial cooling around the top boundary as:
 \begin{eqnarray}
  Q_\mathrm{rad} &=& \frac{1}{r^2}\frac{d}{d r}
   \left[
    r^2\kappa_\mathrm{r}\rho_0 c_\mathrm{p} \frac{dT_0}{dr}
    - r^2F_\mathrm{s}(r)
			  \right],\\
  4\pi r^2F_\mathrm{s} &=& L_\odot \exp\left[-\left(
					     \frac{r-r_\mathrm{max}}{d_\mathrm{s}}
					    \right)^2\right],
 \end{eqnarray}
 where $F_\mathrm{s}$ is the artificial energy flux for the cooling
 around the top boundary with $d_\mathrm{s}=18.8\ \mathrm{Mm}$, which is
 two times the local pressure scale height at the top
 boundary. $\kappa_\mathrm{r}$ is the radiative diffusivity calculated
 with the OPAL repository;
 $L_\odot=3.84\times10^{33}\ \mathrm{erg\ s^{-1}}$ is the solar luminosity.
 \par
 The radial calculation domain extends
 from $r_\mathrm{min}=0.71R_\odot$ to $r_\mathrm{max}=0.96R_\odot$,
 where $R_\odot=6.96\times10^{10}\ \mathrm{cm}$ is the solar radius. The
 whole sphere is covered with the Yin-Yang grid
 \citep{2004GGG.....5.9005K}. We prepare the number of grid points of
 $96(r)\times384(\theta)\times1152(\phi)\times2(\mathrm{Yin-Yang})$.
 In the analyses, we convert the Yin-Yang grid to the almost equivalent
 ordinary spherical geometry with the number of grid points as
 $96(r)\times768(\theta)\times1536(\phi)$.
 Because the near surface layer is not included, we cannot discuss the
 near surface shear layer in this study.
 In this letter, we show two
 cases: HD without the magnetic field and MHD with the magnetic field. We
 add small random perturbation on the entropy for both cases as an
 initial condition. For the case MHD, an axisymmetric longitudinal magnetic
 field $|B_\phi|=100\ \mathrm{G}$ is imposed. The magnetic field is antisymmetric
 about the equator. We calculate the cases HD and MHD for 7000 and 14000
 days, respectively.\par
 We solve the
 equation with the fourth-order space-centered derivative and the four-step
 Runge-Kutta method \citep{2005A&A...429..335V}. An artificial
 viscosity suggested by \cite{2014ApJ...789..132R} is also adopted to
 stabilize the numerical calculation. We do not adopt any explicit
 diffusivity to maximize the resolution.\par
 Averages for the analyses are done in 5000-7000 and
 10000-14000 days for the cases HD and MHD, respectively.
 
 \section{Results}
\begin{figure}[htb!]
   \centering
   \includegraphics[width=1.\textwidth]{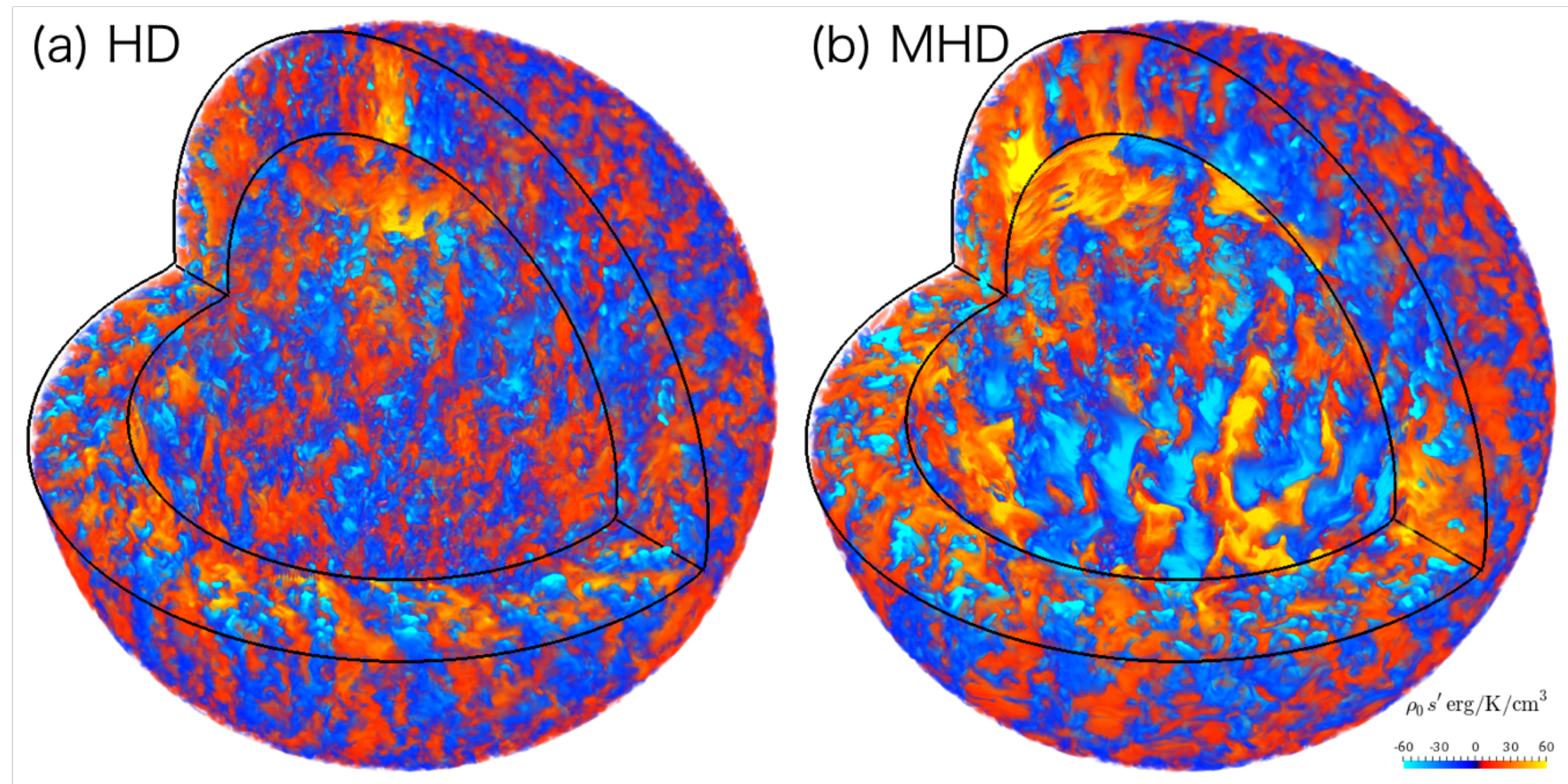}
   \caption{ 3D volume rendering of the perturbation of
   the specific entropy multiplied by the background density for the
   cases (a) HD and (b) MHD are shown. A corresponding movie for the
   panel B is available online.
   The animation continues seven seconds with covering
   40 days of the global convection. The turbulent nature of the thermal
   convection is clearly seen in the animation.
   \label{3d}}
  \end{figure}

 The overall convection structure in the cases HD (panel a) and MHD
 (panel b) is shown in Fig. \ref{3d}.
 The perturbation of the entropy
 $s' = s - \langle s\rangle$ multiplied by the background density is
 shown.
 Compared with the result in \cite{2016Sci....351..1427}, in
 which the thermal conductivity on the entropy is significantly large,
 we can observe the small-scale entropy structure in Fig. \ref{3d}.
 A rotationally
 aligned structure as a result of the Coriolis force is also observed (see also
 the corresponding movie online.). Because of the high resolution, the
 small-scale dynamo is well excited. We confirm that the
 small-scale magnetic energy exceeds the kinetic energy at a small scale
 $\ell>200$, where $\ell$ is the spherical harmonic degree. The turbulent
 magnetic energy around the base of the convection zone ($<0.8R_\odot$)
 exceeds the kinetic energy. These facts indicate that the flow is
 significantly affected by the small-scale magnetic field.
 This is seen in the comparison between the panels a and
 b in Fig. \ref{3d}.
 During the average period (10000-14000 day in the case MHD), the
 large-scale magnetic field does not show any polarity reversal. 
 We note that the polarity reversal is seen around 4500 and 8000 day.
 \par
\begin{figure}[htb!]
   \centering
   \includegraphics[width=0.8\textwidth]{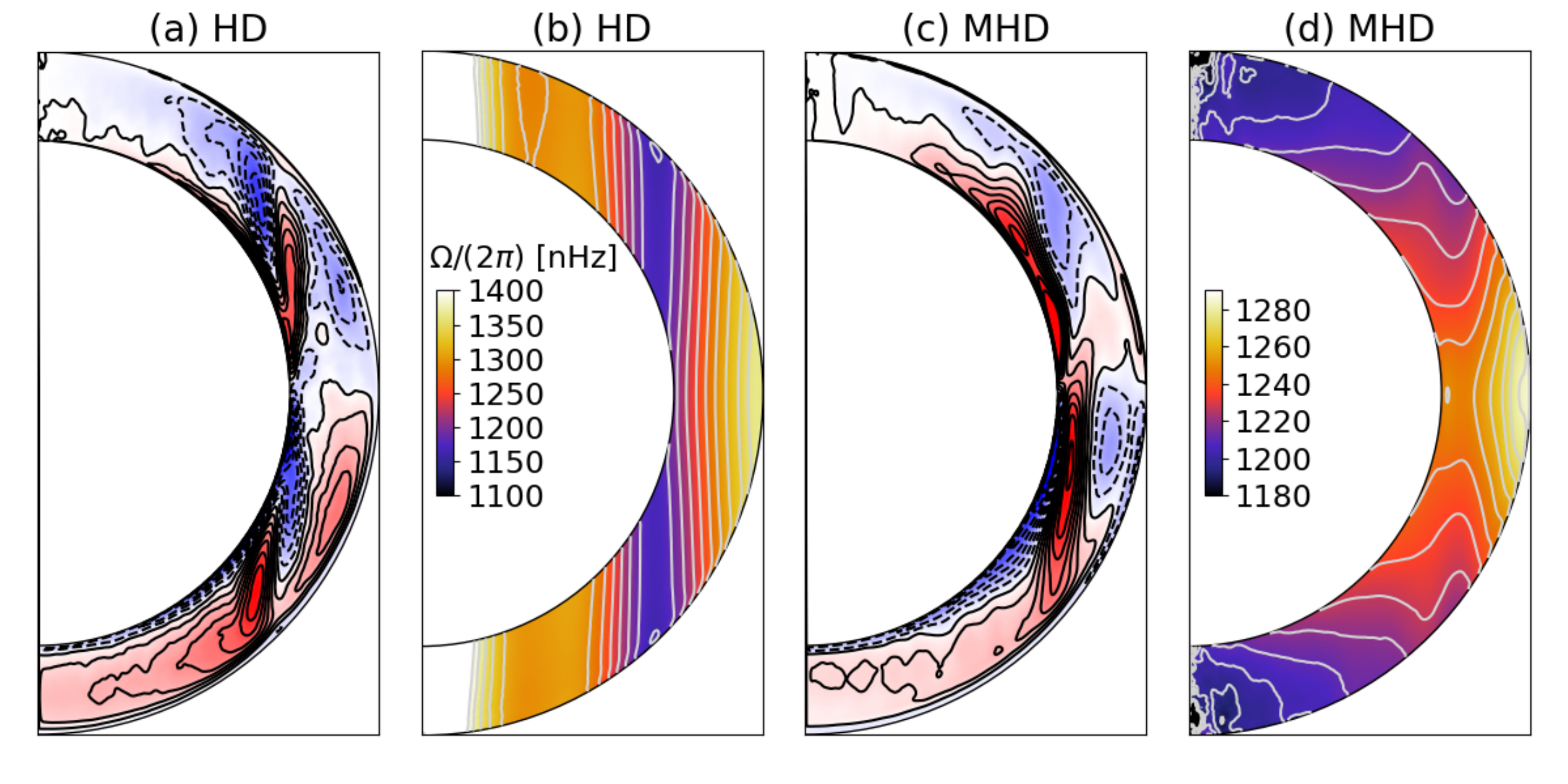}
   \caption{\label{om}
   The panels a and c show the streamline of the meridional flow for
   the cases HD and MHD, respectively. The solid and dashed lines show
   the clockwise and anticlockwise flows, respectively.
   The angular velocity $\Omega/(2\pi)$ is shown
   for the cases HD and MHD in panels b and d, respectively, where
   $\Omega = v_\phi/\lambda+\Omega_0$.
   }
  \end{figure}

   Figs. \ref{om}a and c show the streamline of the
   mass flux of the meridional flow
   $\rho_0 \langle{\bs v}_\mathrm{m}\rangle$ for the cases HD and MHD,
   respectively, where
   ${\bs v}_\mathrm{m}$ includes the radial and latitudinal components
   of the velocity.
   The
   meridional flow in the case HD (Fig. \ref{om}a) shows a north-south
   aligned feature around the tangential cylinder. The overall flow
   pattern is anticlockwise (clockwise) in the northern (southern)
   hemisphere. In the case MHD (Fig. \ref{om}c), we see a clockwise
   (anticlockwise)
   meridional flow cell around the base of the convection zone in the
   northern (southern) hemisphere.   
   Figs. \ref{om}b and d show the angular velocity for the cases HD
   and MHD, respectively.
   Because of the strong Lorentz force of the
   dynamo-generated magnetic field, the shear of the differential
   rotation, i.e. $\Delta \Omega$, decreased in the case MHD (see the
   difference in the color bar). In addition, the polar region is
   accelerated as well as the equator in the case HD. This feature is 
   frequently seen in hydrodynamic
   calculations \citep{2000ApJ...532..593M,2012A&A...546A..19G}. The
   most important difference between the cases HD and MHD concerns the
   Taylor-Proudman balance. In the case HD, the contour lines of the
   angular velocity are almost parallel to the rotational axis (Fig. \ref{om}b). This
   indicates that the case HD is in the Taylor-Proudman state.
   %, in which
   %the Coriolis force term
   %($-\lambda\partial\langle\Omega\rangle^2/\partial z$) in
   %eq. (\ref{tp})
   %is dominant and no other term
   %is balanced with this.
   In contrast, the angular velocity in the
   case MHD shows significant deviation from the Taylor-Proudman
   state. This indicates that some terms in eq. (\ref{tp}) are balanced with the
   Coriolis force term.
   \par
\begin{figure}[htb!]
   \centering
   \includegraphics[width=0.8\textwidth]{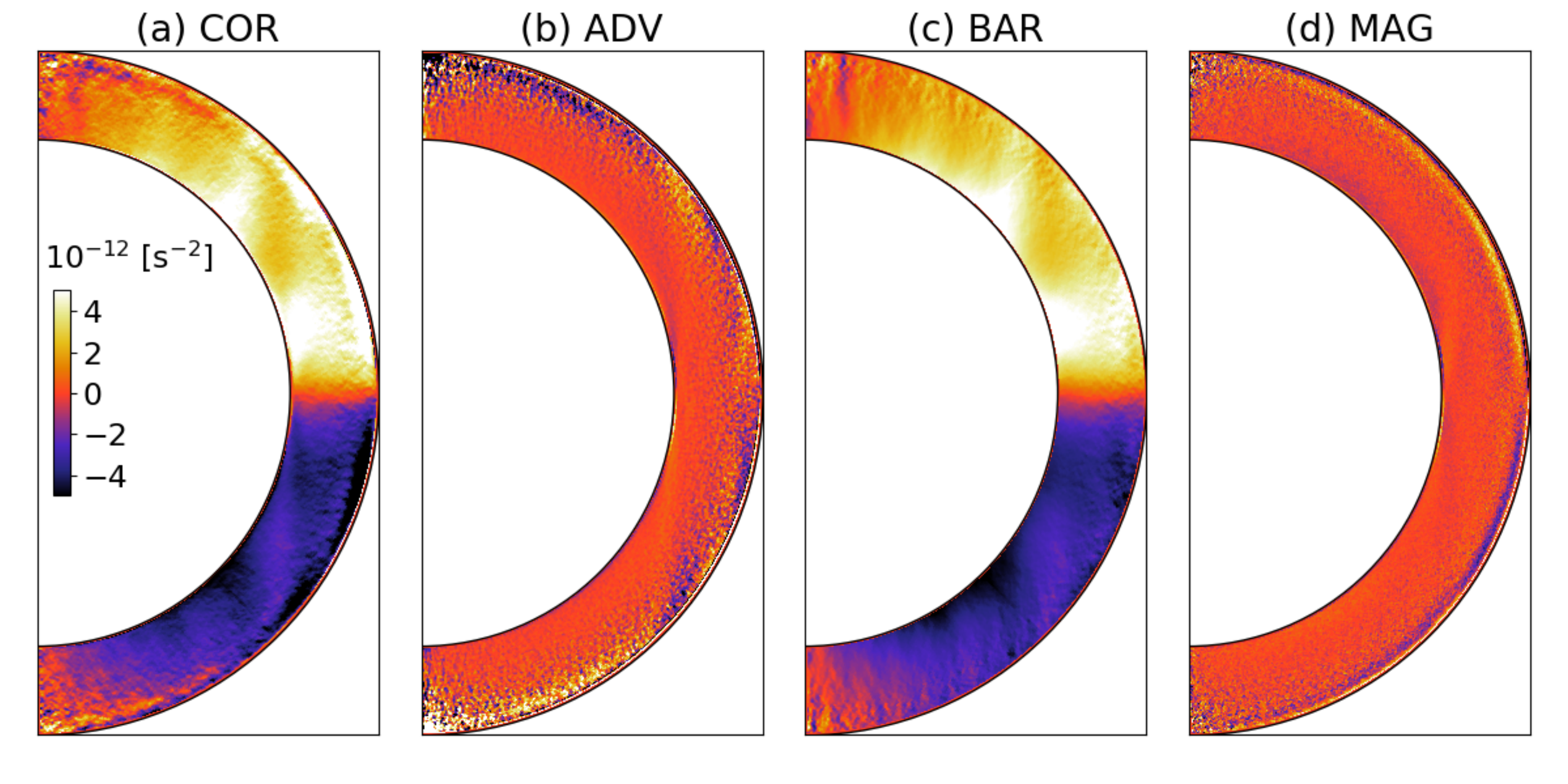}
   \caption{\label{balance} Each term in eq. (\ref{eq_balance}) in the
   case MHD is shown.}
  \end{figure}

   To investigate the force balance on the meridional plane, we
   transpose eq. (\ref{tp}) as:
   \begin{eqnarray}
    \mathrm{COR} &=& \mathrm{ADV} + \mathrm{BAR} + \mathrm{MAG},\label{eq_balance}\\
     \mathrm{COR} &=& -\lambda
      \frac{\partial\langle\Omega\rangle^2}{\partial z},\\
    \mathrm{ADV} &=&
     \left(\nabla\times\langle\bs{v}\times\bs{\omega}\rangle\right)_\phi,\\
    \mathrm{BAR} &=&
     -\frac{g}{c_\mathrm{p}r}\frac{\partial\langle s\rangle}{\partial
     \theta},\\
    \mathrm{MAG} &=&
     \left(
      \nabla\times
      \left\langle
       \frac{1}{4\pi\rho}(\nabla\times{\bs B})\times{\bs B}
      \right\rangle
     \right)_\phi,
   \end{eqnarray}
   where $\partial \langle \omega_\phi\rangle/\partial t$ is ignored,
   because the calculation is already in a statistically steady state.   
   Fig. \ref{balance} shows the balance in the case MHD. Minor
   contributions from
   ADV and MAG around the top boundary are seen in Fig. \ref{balance}b
   and d, but the Coriolis force (COR: Fig. \ref{balance}a) is
   mainly balanced with the
   baroclinic term; i.e., the entropy gradient (BAR:
   Fig. \ref{balance}c). Although the difference in differential rotation is
   caused by including the magnetic field, the Lorentz force MAG is
   not the main contribution to the difference.
  \par
\begin{figure}[htb!]
   \centering
   \includegraphics[width=0.8\textwidth]{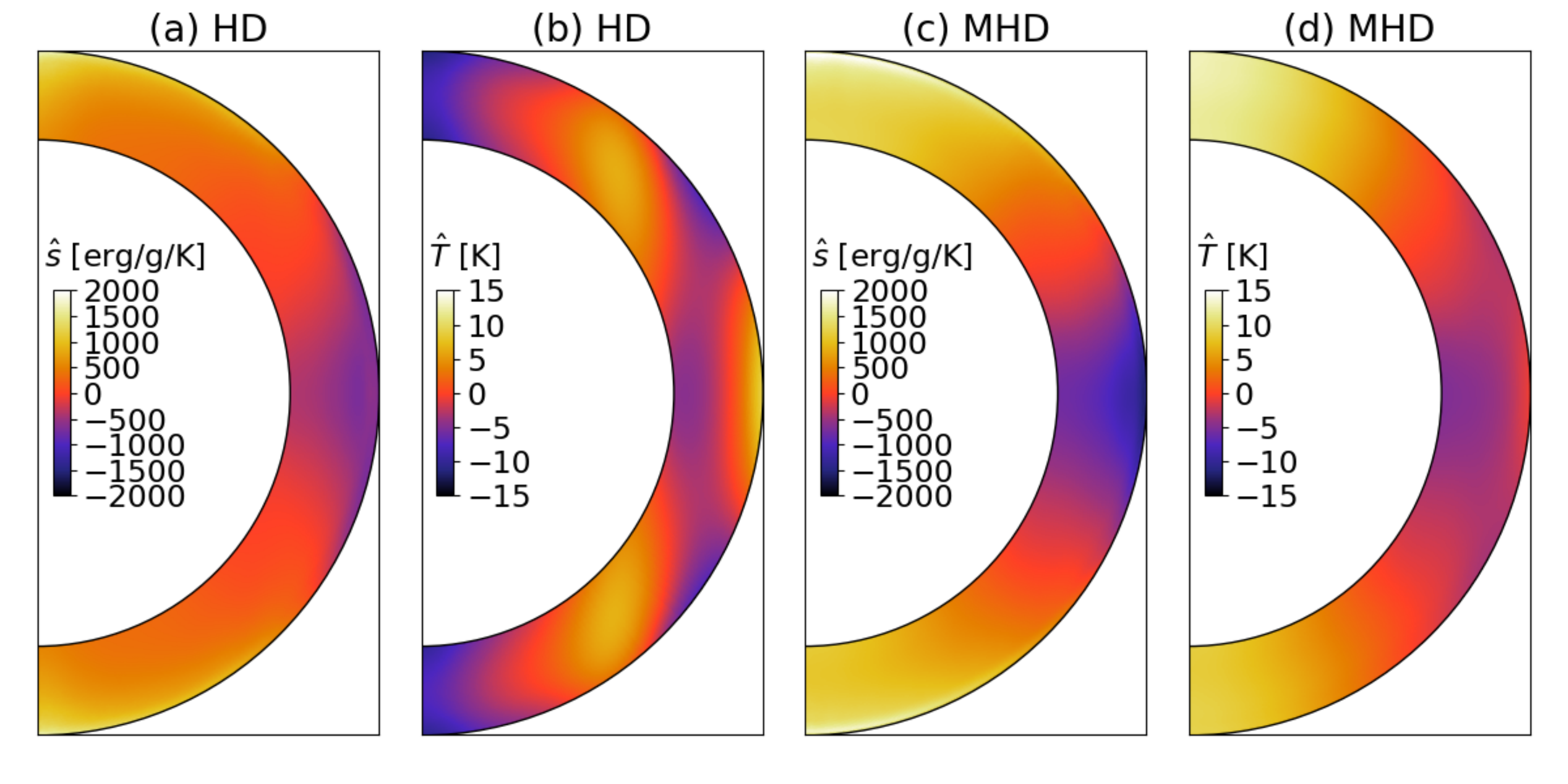}
   \caption{\label{sete}. Latitudinal distributions of the entropy
   $\hat{s}$ and the temperature $\hat{T}$ are shown in the cases HD and
   MHD. The panels a and b show the result of the case HD, and c and d
   show the result of the case MHD.  The entropy $\hat{s}$ (temperature
   $\hat{T}$) is shown in the panels a and c (b and d).}
  \end{figure}
\begin{figure}[htb!]
   \centering
   \includegraphics[width=0.8\textwidth]{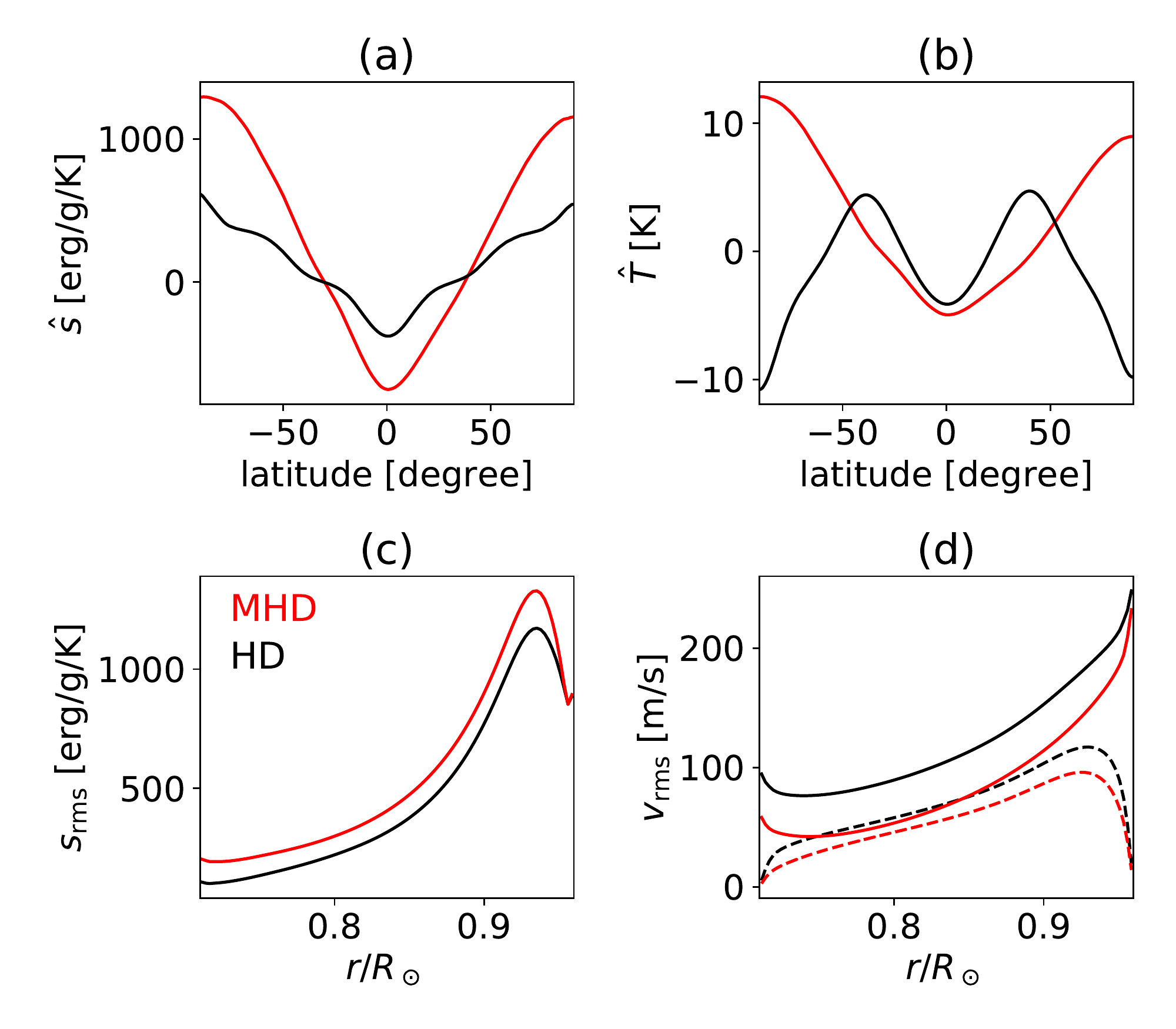}
   \caption{\label{comp} The latitudinal distributions of the entropy
   ($\hat{s}$, panel a) and the temperature ($\hat{T}$, panel b) at the
   bottom boundary are shown. Panel c shows the RMS
   entropy perturbation $s_\mathrm{rms}$. The panel d shows the RMS
   convection velocity.
   The dashed and solid lines show the radial and horizontal components.
   Black and red lines show the results of the
   cases HD and MHD, respectively.}
  \end{figure}

  Fig. \ref{sete} shows the latitudinal distributions of the entropy and
  the temperature in the cases HD and MHD. The hat for a value $Q$ is
  expressed as:
  \begin{eqnarray}
   \hat{Q} = \langle Q\rangle - \bar{Q},
  \end{eqnarray}
  where $\bar{Q}$ is the horizontal average of the value
  $Q$. The value $\hat{Q}$ is useful to show the
  latitudinal distribution with dismissing the radial distribution.
  Fig. \ref{sete}a and c show $\hat{s}$ in the cases HD and MHD,
  respectively. It is clear that the latitudinal entropy gradient is
  increased throughout the convection zone in the case MHD.
  The reasons for this are explained in the following paragraph.
  For
  reference, the latitudinal temperature distributions $\hat{T}$ of the 
  cases HD and MHD are shown in Fig. \ref{sete}b and d, respectively. In the
  case HD, the temperature decreases to the pole from the middle latitude, whereas
  in the case MHD, the temperature
  increases monotonically to the pole. This is also
  shown in Figs. \ref{comp}a and b as a one-dimensional plot at the
  base of the convection zone.
  \par
  There are two reasons that the latitudinal entropy gradient is increased
  in the case MHD. One is that the perturbation of the entropy
  is increased in this case. Fig. \ref{comp}c shows the
  root-mean-square (RMS) entropy in the cases HD (black) and MHD (red).
  It is clear that the RMS entropy in the case MHD is increased compared
  with the case HD.
  This effect is already
  reported in the high-resolution calculation of the small-scale dynamo in
  a nonrotating Cartesian geometry \citep{2015ApJ...803...42H}. When
  the
  turbulent velocity between
  the up- and downflows is suppressed by the small-scale
  magnetic field, the amplitude of the entropy perturbation is
  increased. This increases the anisotropy of the latitudinal energy
  flux, i.e, the CZ effect. The other reason is the suppression of the
  convection
  velocity.
  Fig. \ref{comp}d shows the RMS velocity in
  the cases HD (black) and MHD (red). The dashed and solid lines show
  radial and horizontal component, respectively.
  The convection velocity in the
  case MHD
  is clearly suppressed because of the Lorentz
  force. This indicates that the Coriolis force becomes more effective on the slow
  convection in the case MHD. This also increases the anisotropy of the
  latitudinal energy flux.
  This is confirmed with estimating the correlation of entropy
  perturbation and the latitudinal velocity $\langle
  s'v_\theta\rangle$, which is proportional to the latitudinal enthalpy flux.
  The average is done in the initial phase of the calculation (50-500
  day) where the latitudinal entropy gradient is not well established
  and the contribution from the isotropic turbulent thermal conduction
  is small.
  We can see poleward anisotropic energy flux in almost all
  locations in both cases. In the case HD, we see equatorward energy
  flux in the small
  region close to the equator and the base of the convection zone.
  We summarize the reasons. The entropy in the upflow (downflow) becomes hotter
  (cooler) due to the magnetic field.
  There are several possible influences by the magnetic field to
  increases the entropy perturbation. While the magnetic field decreases
  the convection velocity, the total energy flux does not change and the
  entropy perturbation increases to compensate the slow convection
  velocity.
  In addition, the time for the cooling around the
  top boundary becomes long due to slow convection velocity with the
  magnetic field. This also increase the entropy perturbation. Since the
  magnetic field becomes strong and the convective motion is
  significantly affected, the entropy perturbation and convection
  velocity is decoupled. This causes a deviation from the classical mixing
  length theory.
  Then, the hot upflow (cool
  downflow) is bent poleward (equatorward) by the Coriolis force. Since
  the convection velocity is suppressed this second step is also
  promoted by the magnetic field. As a result the latitudinal entropy
  gradient (hot pole and cool equator) is amplified by the magnetic
  field.
  \par
  Regarding the deviation from the
  Taylor-Proudman state, the Lorentz force on the differential
  rotation also has a role. As explained, the Lorentz force is strong enough
  to reduce the shear of the differential rotation $\Delta
  \Omega$. We transform eq. (\ref{eq_balance}) by including only
  the terms COR and BAR as:
  \begin{eqnarray}
   -2\lambda\langle\Omega\rangle\frac{\partial \langle\Omega_1\rangle}{\partial z}
    = -\frac{g}{c_\mathrm{p}r}\frac{\partial\langle s\rangle}{\partial\theta},
  \end{eqnarray}
  where $\Omega_1 = v_\phi/\lambda$. This relation shows that even
  with the same latitudinal entropy gradient, the Taylor-Proudman state
  is more easily broken with a small shear of the angular velocity $\langle
  \Omega_1\rangle$. Thus, the Lorentz force on the differential rotation
  also promotes the deviation of the Taylor-Proudman state.
  
  \section{Summary}
  In this paper, we carry out a relatively high-resolution calculation of
 the stellar global convection in the spherical shell. We investigate the
 effect of the efficient small-scale dynamo on the shape of the
 differential rotation. Compared with our previous calculation
 \citep{2016Sci....351..1427}, we exclude the strong thermal conductivity on
 the entropy in order not to suppress the small-scale thermal
 features. Here we compare the calculations with and without the
 magnetic field. Because of the relatively high resolution, the efficient
 small-scale dynamo is achieved, even covering the whole convection
 zone.\par
 One important finding is that the differential rotation deviates
 significantly from the Taylor-Proudman state when the
 magnetic field is included. The reasons for this
 deviation are summarized as follows.
 \begin{enumerate}
  \item The Lorentz force suppresses the differential rotation. Thus,
	the criterion, i.e., required latitudinal entropy gradient to break the
	Taylor-Proudman balance, is relaxed.
  \item The entropy perturbation is increased by suppressing
	turbulent velocity
	between the up- and downflows by the small-scale magnetic
	field. This increases the anisotropy of the latitudinal energy
	flux.	
  \item The convection velocity is suppressed by the magnetic
	field. Thus, the rotational effect is increased in the MHD
	calculation. This also increases the anisotropy of the
	latitudinal energy flux.
 \end{enumerate}
 Reason 3 above possibly relates to reason 1. When the convection
 velocity is suppressed, the angular momentum transport by the
 convection tends to be suppressed, leading to the reduction in the
 differential rotation.
 Reasons 2 and 3 increase the latitudinal entropy gradient that
 breaks the Taylor-Proudman balance.
 Reasons 2 and 3 are possibly achieved with a high effective thermal
 Prandtl number, in which the thermal conductivity is low and the
 kinetic viscosity is strong. There are several attempts at mimicking
 the small-scale dynamo and magnetic field by adopting a high
 effective Prandtl number
 \citep{2016AdSpR..58.1475O,2017ApJ...851...74B,2018arXiv180100560K}.
 This type of attempt is valuable, because the small-scale dynamo that
 requires a high numerical cost can be reproduced with a reasonable
 cost.
 To mimic the small-scale dynamo correctly, detailed comparisons
 between the calculation in this study and calculations with high
 Prandtl numbers are required. 
 In \cite{2018arXiv180100560K}, the generation of the
 negative latitudinal entropy gradient in their high Prandtl number
 calculations is discussed. They suggest that the entropy gradient is caused by the
 interaction of the meridional flow and the subadiabatic layer around
 the base of the convection zone. In this calculation, the subadiabatic
 layer around the base of the convection zone in the case MHD is
 confirmed, whereas the case HD does not have such a layer. The meridional
 flow, however, in this calculation, is clockwise (anticlockwise) around
 the base of the convection zone. This tends to generate the positive (negative)
 entropy gradient in the northern (southern) hemisphere. Thus, we
 conclude that the interaction between the
 meridional flow and the subadiabatic layer does not contribute to the
 deviation of the Taylor-Proudman state in this study.
 \par
 In this study, we do not include the radiation zone
 \citep{2011ApJ...742...79B}, which also tends to amplify the
 latitudinal entropy gradient. In addition, the property of the
 small-scale dynamo is affected by the radiation zone
 \citep{2017ApJ...843...52K}. A combination of the small-scale dynamo and
 the radiation zone would determine the detailed shape of the stellar
 differential rotation. This will be addressed in the future.

 \acknowledgements
 The author would like to thank the anonymous referee for his/her great suggestions.
 The author is grateful to Y. Bekki for his insightful comments on the manuscript.
 These results were obtained by using the K computer at the RIKEN Advanced
 Institute for Computational Science (Proposal number hp180042, hp170239, hp170012, hp160026, hp160252,
 ra000008). This work was supported by MEXT/JSPS KAKENHI Grant Number
 JP18H04436, JP16K17655, JP16H01169.
 This research was supported by MEXT as ``Exploratory Challenge on Post-K
 computer'' (Elucidation of the Birth of Exoplanets [Second Earth] and
 the Environmental Variations of Planets in the Solar System)

%\bibliographystyle{apj}
%\bibliography{apj-jour,reference}

\begin{thebibliography}{27}
\expandafter\ifx\csname natexlab\endcsname\relax\def\natexlab#1{#1}\fi

\bibitem[{{Bekki} {et~al.}(2017){Bekki}, {Hotta}, \&
  {Yokoyama}}]{2017ApJ...851...74B}
{Bekki}, Y., {Hotta}, H., \& {Yokoyama}, T. 2017, \apj, 851, 74

\bibitem[{{Brun} {et~al.}(2011){Brun}, {Miesch}, \&
  {Toomre}}]{2011ApJ...742...79B}
{Brun}, A.~S., {Miesch}, M.~S., \& {Toomre}, J. 2011, \apj, 742, 79

\bibitem[{{Christensen-Dalsgaard} {et~al.}(1996){Christensen-Dalsgaard},
  {Dappen}, {Ajukov}, {Anderson}, {Antia}, {Basu}, {Baturin}, {Berthomieu},
  {Chaboyer}, {Chitre}, {Cox}, {Demarque}, {Donatowicz}, {Dziembowski},
  {Gabriel}, {Gough}, {Guenther}, {Guzik}, {Harvey}, {Hill}, {Houdek},
  {Iglesias}, {Kosovichev}, {Leibacher}, {Morel}, {Proffitt}, {Provost},
  {Reiter}, {Rhodes}, {Rogers}, {Roxburgh}, {Thompson}, \&
  {Ulrich}}]{1996Sci...272.1286C}
{Christensen-Dalsgaard}, J., {et~al.} 1996, Science, 272, 1286

\bibitem[{{Fan} \& {Fang}(2014)}]{2014ApJ...789...35F}
{Fan}, Y., \& {Fang}, F. 2014, \apj, 789, 35

\bibitem[{{Gastine} {et~al.}(2012){Gastine}, {Duarte}, \&
  {Wicht}}]{2012A&A...546A..19G}
{Gastine}, T., {Duarte}, L., \& {Wicht}, J. 2012, \aap, 546, A19

\bibitem[{{Gastine} {et~al.}(2014){Gastine}, {Yadav}, {Morin}, {Reiners}, \&
  {Wicht}}]{2014MNRAS.438L..76G}
{Gastine}, T., {Yadav}, R.~K., {Morin}, J., {Reiners}, A., \& {Wicht}, J. 2014,
  \mnras, 438, L76

\bibitem[{{Hotta}(2017)}]{2017ApJ...843...52K}
{Hotta}, H. 2017, \apj, 52

\bibitem[{{Hotta} {et~al.}(2014){Hotta}, {Rempel}, \&
  {Yokoyama}}]{2014ApJ...786...24H}
{Hotta}, H., {Rempel}, M., \& {Yokoyama}, T. 2014, \apj, 786, 24

\bibitem[{{Hotta} {et~al.}(2015{\natexlab{a}}){Hotta}, {Rempel}, \&
  {Yokoyama}}]{2015ApJ...803...42H}
---. 2015{\natexlab{a}}, \apj, 803, 42

\bibitem[{{Hotta} {et~al.}(2015{\natexlab{b}}){Hotta}, {Rempel}, \&
  {Yokoyama}}]{2015ApJ...798...51H}
---. 2015{\natexlab{b}}, \apj, 798, 51

\bibitem[{{Hotta} {et~al.}(2016){Hotta}, {Rempel}, \&
  {Yokoyama}}]{2016Sci....351..1427}
---. 2016, Science, 351, 1427

\bibitem[{{Hotta} {et~al.}(2012){Hotta}, {Rempel}, {Yokoyama}, {Iida}, \&
  {Fan}}]{2012A&A...539A..30H}
{Hotta}, H., {Rempel}, M., {Yokoyama}, T., {Iida}, Y., \& {Fan}, Y. 2012, \aap,
  539, A30

\bibitem[{{Hotta} \& {Yokoyama}(2011)}]{2011ApJ...740...12H}
{Hotta}, H., \& {Yokoyama}, T. 2011, \apj, 740, 12

\bibitem[{{Kageyama} \& {Sato}(2004)}]{2004GGG.....5.9005K}
{Kageyama}, A., \& {Sato}, T. 2004, Geochemistry, Geophysics, Geosystems, 5,
  9005

\bibitem[{{K{\"a}pyl{\"a}} {et~al.}(2014){K{\"a}pyl{\"a}}, {K{\"a}pyl{\"a}}, \&
  {Brandenburg}}]{2014A&A...570A..43K}
{K{\"a}pyl{\"a}}, P.~J., {K{\"a}pyl{\"a}}, M.~J., \& {Brandenburg}, A. 2014,
  \aap, 570, A43

\bibitem[{{K{\"a}pyl{\"a}} {et~al.}(2018){K{\"a}pyl{\"a}}, {K{\"a}pyl{\"a}}, \&
  {Brandenburg}}]{2018arXiv180209607K}
---. 2018, ArXiv e-prints

\bibitem[{{K{\"a}pyl{\"a}} {et~al.}(2017){K{\"a}pyl{\"a}}, {K{\"a}pyl{\"a}},
  {Olspert}, {Warnecke}, \& {Brandenburg}}]{2017A&A...599A...4K}
{K{\"a}pyl{\"a}}, P.~J., {K{\"a}pyl{\"a}}, M.~J., {Olspert}, N., {Warnecke},
  J., \& {Brandenburg}, A. 2017, \aap, 599, A4

\bibitem[{{Karak} {et~al.}(2018){Karak}, {Miesch}, \&
  {Bekki}}]{2018arXiv180100560K}
{Karak}, B.~B., {Miesch}, M., \& {Bekki}, Y. 2018, ArXiv e-prints

\bibitem[{{Miesch} {et~al.}(2006){Miesch}, {Brun}, \&
  {Toomre}}]{2006ApJ...641..618M}
{Miesch}, M.~S., {Brun}, A.~S., \& {Toomre}, J. 2006, \apj, 641, 618

\bibitem[{{Miesch} {et~al.}(2000){Miesch}, {Elliott}, {Toomre}, {Clune},
  {Glatzmaier}, \& {Gilman}}]{2000ApJ...532..593M}
{Miesch}, M.~S., {Elliott}, J.~R., {Toomre}, J., {Clune}, T.~L., {Glatzmaier},
  G.~A., \& {Gilman}, P.~A. 2000, \apj, 532, 593

\bibitem[{{O'Mara} {et~al.}(2016){O'Mara}, {Miesch}, {Featherstone}, \&
  {Augustson}}]{2016AdSpR..58.1475O}
{O'Mara}, B., {Miesch}, M.~S., {Featherstone}, N.~A., \& {Augustson}, K.~C.
  2016, Advances in Space Research, 58, 1475

\bibitem[{{Parker}(1955)}]{1955ApJ...122..293P}
{Parker}, E.~N. 1955, \apj, 122, 293

\bibitem[{{Rempel}(2005)}]{2005ApJ...622.1320R}
{Rempel}, M. 2005, \apj, 622, 1320

\bibitem[{{Rempel}(2014)}]{2014ApJ...789..132R}
---. 2014, \apj, 789, 132

\bibitem[{{Rogers} {et~al.}(1996){Rogers}, {Swenson}, \&
  {Iglesias}}]{1996ApJ...456..902R}
{Rogers}, F.~J., {Swenson}, F.~J., \& {Iglesias}, C.~A. 1996, \apj, 456, 902

\bibitem[{{Schou} {et~al.}(1998){Schou}, {Antia}, {Basu}, {Bogart}, {Bush},
  {Chitre}, {Christensen-Dalsgaard}, {di Mauro}, {Dziembowski}, {Eff-Darwich},
  {Gough}, {Haber}, {Hoeksema}, {Howe}, {Korzennik}, {Kosovichev}, {Larsen},
  {Pijpers}, {Scherrer}, {Sekii}, {Tarbell}, {Title}, {Thompson}, \&
  {Toomre}}]{1998ApJ...505..390S}
{Schou}, J., {et~al.} 1998, \apj, 505, 390

\bibitem[{{V{\"o}gler} {et~al.}(2005){V{\"o}gler}, {Shelyag}, {Sch{\"u}ssler},
  {Cattaneo}, {Emonet}, \& {Linde}}]{2005A&A...429..335V}
{V{\"o}gler}, A., {Shelyag}, S., {Sch{\"u}ssler}, M., {Cattaneo}, F., {Emonet},
  T., \& {Linde}, T. 2005, \aap, 429, 335

\end{thebibliography}

\end{document}